\documentclass[twocolumn]{aastex631}
\usepackage{CJK}
\usepackage{soul}

\shorttitle{Inferring flows from magnetic fields}
\shortauthors{Dash et al.}

\begin{document}
\begin{CJK*}{UTF8}{gbsn}
\title{Ensemble Kalman Filter Data Assimilation Into Surface Flux Transport Model To Infer Surface Flows: An Observing System Simulation Experiment}

\correspondingauthor{Soumyaranjan Dash}
\email{dashs@hawaii.edu}

\author[0000-0003-0103-6569]{Soumyaranjan Dash}
\affiliation{Institute for Astronomy, University of Hawai`i at M\={a}noa, Pukalani, HI 96768, USA}

\author[0000-0002-6338-0691]{Marc L. DeRosa}
\affiliation{Lockheed Martin Solar and Astrophysics Laboratory, Palo Alto, CA 94306, USA}

\author[0000-0002-2227-0488]{Mausumi Dikpati}
\affiliation{High Altitude Observatory, NSF-NCAR, 3080 Center Green Drive, Boulder, CO 80301, USA}

\author[0000-0003-4043-616X]{Xudong Sun (孙旭东)}
\affiliation{Institute for Astronomy, University of Hawai`i at M\={a}noa, Pukalani, HI 96768, USA}

\author[0000-0003-1753-8002]{Sushant S. Mahajan}
\affiliation{W.W. Hansen Experimental Physics Laboratory, Stanford University, Stanford, CA 94305-4085, USA}

\author[0000-0002-0671-689X]{Yang Liu (刘扬)}
\affiliation{W.W. Hansen Experimental Physics Laboratory, Stanford University, Stanford, CA 94305-4085, USA}

\author[0000-0001-9130-7312]{J. Todd Hoeksema}
\affiliation{W.W. Hansen Experimental Physics Laboratory, Stanford University, Stanford, CA 94305-4085, USA}

\begin{abstract}

Knowledge of the global magnetic field distribution and its evolution on the Sun's surface is crucial for modeling the coronal magnetic field, understanding solar wind dynamics, computing the heliospheric open flux distribution and predicting solar cycle strength. As the far side of the Sun cannot be observed directly and high-latitude observations always suffer from projection effects, we often rely on surface flux transport simulations (SFT) to model long-term global magnetic field distribution. Meridional circulation, the large-scale north-south component of the surface flow profile, is one of the key components of the SFT simulation that requires further constraints near high latitudes. Prediction of the photospheric magnetic field distribution requires knowledge of the flow profile in the future, which demands reconstruction of that same flow at the current time so that it can be estimated at a later time. By performing Observing System Simulation Experiments, we demonstrate how the Ensemble Kalman Filter technique, when used with a SFT model, can be utilized to make ``posterior'' estimates of flow profiles into the future that can be used to drive the model forward to forecast photospheric magnetic field distribution.

\end{abstract}


\section{Introduction} \label{sec:intro}

Magnetic fields on the Sun drive a diverse range of phenomena that span different time scales. Solar flares, coronal mass ejections, and radiation from energetic particles influence our space environment in the short term, whereas emergence of sunspots and magnetic flux transport on the photosphere control the solar magnetic activity cycle that covers a longer time scale. Overall, these events impact Earth's space environment, affecting satellite operations and telecommunication, and contribute to space weather variations. Knowledge of the surface magnetic field distribution and evolution is crucial for predicting both short-term space weather and long-term climate variations. The Sun's surface magnetic field distribution is extremely important and is used in multiple applications as a boundary condition for modelling coronal magnetic fields, initial condition for solar wind models as well as to compute the heliospheric open solar flux distribution. Polar magnetic field evolution also is closely linked to behavior of the solar cycle. It allows us to make predictions about the amplitude and timing of solar cycles \citep{Upton2014ApJ,Bhowmik2018NatCo,Nandy2021SoPh,Jha2024ApJ,Pal2024MNRAS}, which have implications on long-term climate variability.

The solar dynamo is essentially an interplay between the poloidal and toroidal components of the Sun's magnetic field. During a solar activity minimum, the global magnetic field is dominated by the poloidal component. The solar differential rotation (i.e., the longitudinal component of the large-scale flow) stretches this poloidal field longitudinally to generate the toroidal field component in the interior. \citet{Parker1955ApJ} suggests that due to magnetic buoyancy these amplified toroidal flux ropes manifest as bipolar magnetic regions (BMR) on the surface. While the toroidal flux rope rises through the solar convection zone, Coriolis forces typically tilt it such that one of the sunspots emerges leading in longitude and closer to the equator while the other one is behind (following) in longitude and closer to the pole of the respective hemisphere \citep{DSilva1993AA}. Near the Sun's equator, the leading BMR polarities of two opposite hemispheres cancel each other, and the remaining magnetic flux drifts towards the respective poles via large-scale meridional flow (north-south component of the large-scale flow) and diffusion.

In order to study this flux transport process, we require global magnetic field observations spanning all the longitudes and latitudes. However, current ground- and space-based observing facilities are largely restricted to the Sun-Earth line, which limits our observation window to one half of the Sun's surface.  Such observations are not reliable near high latitudes and limb extrema due to high projection effects. To address this issue, we often utilize the surface flux transport (SFT) model, which solves the radial component of the magnetic induction equation (Eq.\ \ref{eq:1}) with imposed large-scale flow parameters, surface magnetic diffusivity, and observed BMR properties (more details in the review paper by \citealp{Yeates2023SSRv}). Some of the SFT models also use data assimilation (DA) to incorporate the observed magnetic flux to generate a global photospheric magnetic field map. In order to accurately model the magnetic flux transport we require observationally constrained description of global flow parameters i.e., differential rotation and meridional circulation as well as magnetic diffusivity. Meridional flow has been observed in the photosphere and within the upper convection zone, spanning from the equatorial region to approximately 60\textdegree\ latitude in each hemisphere \citep{Komm1993SoPh,Hathaway2010Sci,Mahajan2021ApJ}. Nevertheless, comprehensive knowledge regarding flow speed, profile, and temporal fluctuations over multiple solar cycles on the surface and within the deeper convection zone remains elusive based on current observational data owing in part to increased uncertainty near polar latitudes due to projection effects. Recently, \cite{Mahajan2023ApJ} analyzed variations in the global meridional flow pattern for solar cycle 24 using time-distance helioseismology technique utilizing data from Helioseismic Magnetic Imager (HMI) \citep{Scherrer2012SoPh,Schou2012SoPh} on board Solar Dynamics Observatory (SDO) and found that the meridional flow does show a small variation of order 2~m~s$^{-1}$ far away from active regions, and much larger variation close to active regions due to inflows. The study, however, is still limited to 62.5\textdegree\ latitude due to projection effects.

Despite these limitations, SFT modeling has proven remarkably effective in calculating the distribution of photospheric magnetic flux. Both large-scale surface flow components, i.e., differential rotation and meridional circulation, primarily show latitude-dependent variation. SFT simulations are sensitive to parameterized flow profiles. Modelled surface magnetic field distribution can be calibrated to the observations considering the available observations of polar field or its proxy. In order to achieve this, different SFT models consider slightly different yet solar cycle independent meridional flow profiles and magnetic diffusivity to mimic the flux cancellation and transport process. This raises the question, is it possible to infer the parameter value, e.g. peak flow speed of an assumed meridional flow profile, using the magnetic field observations? In this paper we build an Observation System Simulation Experiment (OSSE) system with the SFT model which will guide us further for building a model using observed magnetic fields to infer the different physical variables e.g., flow parameters, magnetic diffusivity, etc.

Ensemble Kalman Filter (EnKF) data assimilation is ideal for our purpose. The EnKF computational technique uses observations sequentially obtained at a current time together with associated ``prior knowledge'' from numerical computation of physical models to provide Bayesian estimates of the ``posterior'' states of the system (model+observations), while accounting for uncertainties in the observations and using physical model-outputs of the evolved states. Although the motivation for the development of such a technique arose from the need for short-to-medium range weather forecasting in geophysics \citep{Evensen1994JGR}, now this technique has been used extensively in many scientific fields such as Earth upper-atmosphere models \citep{Matsuo2013JGRA,Pedatella2013GRL}, weather prediction models \citep{Ha2017MWRv,Hoteit2021BAMS} and ocean circulation models \citep{Chen2022JAMES}. EnKF is used for estimating or constraining otherwise unknown parameters of a physical model, as well as producing better forecasts that are updated through the most recent observations while integrating the model forward in time than deriving from pure model-outputs. The filtering, in brief, is a statistical perspective for obtaining the posterior distributions of a system's states and their uncertainties at the current time based on all accumulated observations so far. The Kalman filter estimates the state of a system using two steps: (i) estimating the state and uncertainty in that state, which are adjusted to newly available observations and (ii) forecasting the updated state and uncertainty by propagating them forward in time. A model, such as Gaussian error, is used to obtain the first-guess of the states, and then from the error covariance matrices of the observations and the theoretically generated observations from the first-guesses of the states the corrections to the first guess of the state are estimated. The process of combining the first guesses in an ensemble of models with observations to derive the new states is essentially the EnKF data assimilation. A detailed review can be found in \cite{Evensen2003OcDyn}.

Numerous studies have explored the application of data assimilation in solar physics \citep{Brun2007AN}. For example, reconstruction of dynamic solar corona using ensemble Kalman filtering and tomography methods with observations \citep{Butala2010SoPh}, forecasting solar cycle behavior by combining real observations with a reduced $\alpha$-$\Omega$ dynamo model \citep{Kitiashvili2008ApJ}, using the four-dimensional variational method (4D-Var) and a cellular-automaton-based avalanche model to predict simulated solar-flare data \citep{Belanger2007SoPh}, generation of ensemble members of photospheric magnetic field distribution with localized EnKF techniques driven by available observations \citep{Arge2010AIPC,Hickmann2015SoPh}. \cite{Jouve2011ApJ} developed a variational data assimilation framework based on a solar $\alpha$-$\Omega$ dynamo model and validated through synthetic data. \cite{Svedin2013MNRAS} showed how a three-dimensional variational (3D-Var) data assimilation methodology can be utilized to obtain accurate model estimates given a set of observations. \cite{Fournier2013GGG} employed the EnKF framework in a three-dimensional, convection-driven geodynamo model, using surface poloidal magnetic fields for full-state estimation. \cite{Dikpati2014GRL,Dikpati2016ApJ} utilized EnKF sequential data assimilation to reconstruct the time variations in meridional flow speed over multiple solar cycles, utilizing poloidal and toroidal magnetic fields as observational data in a two dimensional kinematic Babcok-Leighton solar dynamo model. 

Utilizing EnKF data assimilation with a one-dimensional SFT model, we aim to reconstruct the parameters of the meridional flow profile. With statistical multidimensional regression analysis using a Bayesian approach, we can compare the magnetic field observations to the model-generated magnetic field for a random initial guess of the meridional peak flow speed. Subsequently, the model parameter (peak flow speed) is adjusted to minimize the deviation between the observed and modeled magnetic field, thereby eventually converging at a parameter value inspired by direct observations.

We build a DA experiment with 1D SFT model to study how sensitive the reconstruction of the peak meridional flow speed is to different tunable DA parameters: (a) Magnetic field observations near high latitudes -- how does the error in reconstruction vary with increasing or decreasing the assimilation interval? (b) Possible error associated with the observations -- how well can the model parameter be reconstructed with different values of error associated with the observations? In the following sections describe our SFT model and different cases for reconstruction of peak meridional flow speed. In \S\ref{sec:models}, we explain our SFT model and mathematical formulation of the EnKF methodology, and \S\ref{sec:results} describes our results. A comprehensive discussion and summary of our study is presented in \S\ref{subsec:discussion}. 


\section{Models} \label{sec:models}


\subsection{Surface flux transport model (SFT)}\label{subsec:sft}
The SFT model is based on the idea that radial magnetic flux on the solar surface is carried around by horizontal plasma flows with no back reaction on these flows \citep{Leighton1964ApJ}. This can be modeled by solving the radial component of the magnetic induction equation on the solar surface,
\begin{equation}
\label{eq:1}
    \frac{\partial B_r}{\partial t} + \mathbf{\nabla}\cdot\Big(u_hB_r\Big) = \eta \nabla^2 B_r + S,
\end{equation}

where $\eta$ denotes the effect of supergranular diffusion and $u_h$ describes the imposed horizontal plasma flows i.e., meridional circulation and differential rotation. The supergranular diffusion is assumed to be a constant quantity, hence it is placed outside of the spatial derivative operator. $B_r(s,\phi,t)$ represents the large-scale mean radial magnetic field defined on a sine-latitude ($s$=$\sin\lambda$) and longitude ($\phi$) grid for the time step $t$. The source terms, i.e., BMRs emerging on the solar surface, are denoted by $S(s,\phi,t)$. Since the evolution of the axial dipole moment depends only on longitude-averaged field \citep{DeVore1984SoPh,Cameron2007ApJ,Iijima2017AA,Petrovay2019AA,Yeates2020SoPh}, we simplify our SFT model by averaging the mean field over longitude $\overline B(s,t) = \frac{1}{2\pi}\int^{2\pi}_{0}B_r(s,\phi,t)d\phi$. The simplified SFT model can be described by
\begin{equation}
\label{eq:2}
    \frac{\partial \overline B}{\partial t} = \frac{\partial}{\partial s}\bigg[ \frac{\eta}{R_{\odot}^2}(1 - s^2)\frac{\partial \overline B}{\partial s} - \frac{v_s(s)}{R_{\odot}}\sqrt{1 - s^2}\overline B\bigg],
\end{equation}
where $v_s(s)$ denotes the meridional circulation and $R_{\odot}$ is the solar radius.
Meridional circulation is parameterized as
\begin{equation}
\label{eq:3}
    v_s(s) = D_u s(1-s^2)^{(p/2)}.
\end{equation}
We can compute the axial dipole moment using the longitudinally averaged magnetic field as,

\begin{equation}
\label{eq:dm}
    DM = \frac{3}{2}\int^{1}_{-1} s\overline B(s,dt)ds
\end{equation}

\begin{figure*}[htpb]
    \centering
    \includegraphics[width=\linewidth]{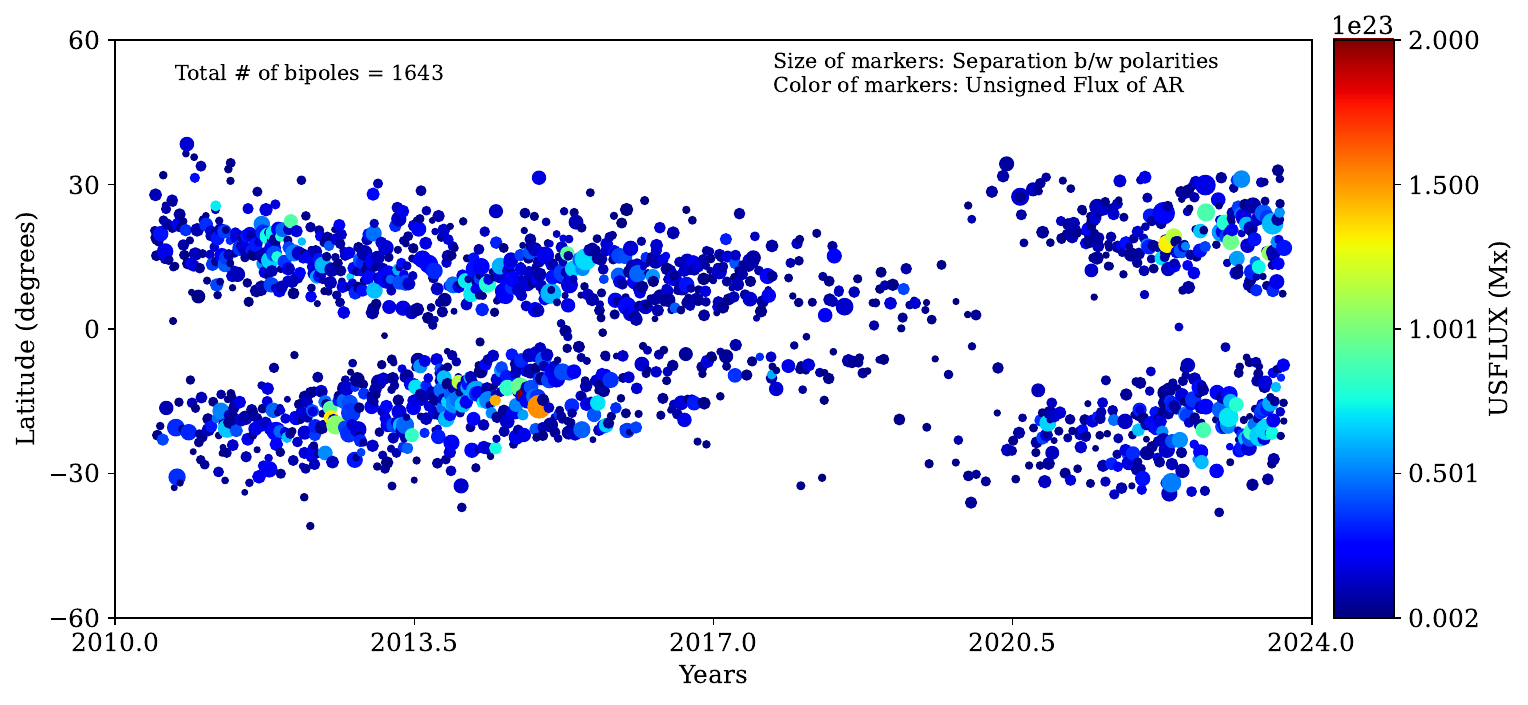}
    \caption{Time-Latitude distribution of BMRs obtained from HMI observations. There are a total of 1643 BMRs identified at their maximum disk-passage area. The color of the markers denotes the unsigned flux and size corresponds to the separation between opposite polarities.}
    \label{fig:0}
\end{figure*}

\cite{Yeates2020SoPh} introduced such a mathematical formulation of the SFT model. We developed our numerical model following \cite{Yeates2020SoPh} in FORTRAN for this study. We solve the model equations numerically using finite-volume method. We adopt their meridional flow profile shape of $p$=2.33, and use $D_u$=0.085~km~s$^{-1}$,  which corresponds to a meridional peak flow speed of $v_0$=30.7~m~s$^{-1}$. In the sine-latitude grid there are 180 grid points considered for our simulation. The source terms are modeled as idealized BMRs using the observed sunspot properties taken from the HMI database spanning a time frame from 17 June 2010 to 04 September 2023 following the formalism provided by \cite{Yeates2020SoPh}. Figure~\ref{fig:0} shows the time-latitude distribution of selected BMRs, where the color indicates the unsigned flux magnitude and the size of the markers denotes the separation between opposite polarities. There are 1643 BMRs identified for incorporation into our SFT model. The BMR properties are extracted from the \texttt{hmi.sharp\_cea\_720s} series available through Joint Science Operations Center (JSOC) using the script distributed by \cite{Yeates2020SoPh}. It is important to note that all the BMR properties are considered at their maximum areal coverage during disk passage. Utilizing these properties we model the source terms for our SFT model following the algorithm described in the same paper. Our SFT model can be accessed at \url{https://github.com/sr-dash/SFT-1D}.

\begin{figure*}[htpb]
    \centering
    \includegraphics[width=0.9\linewidth]{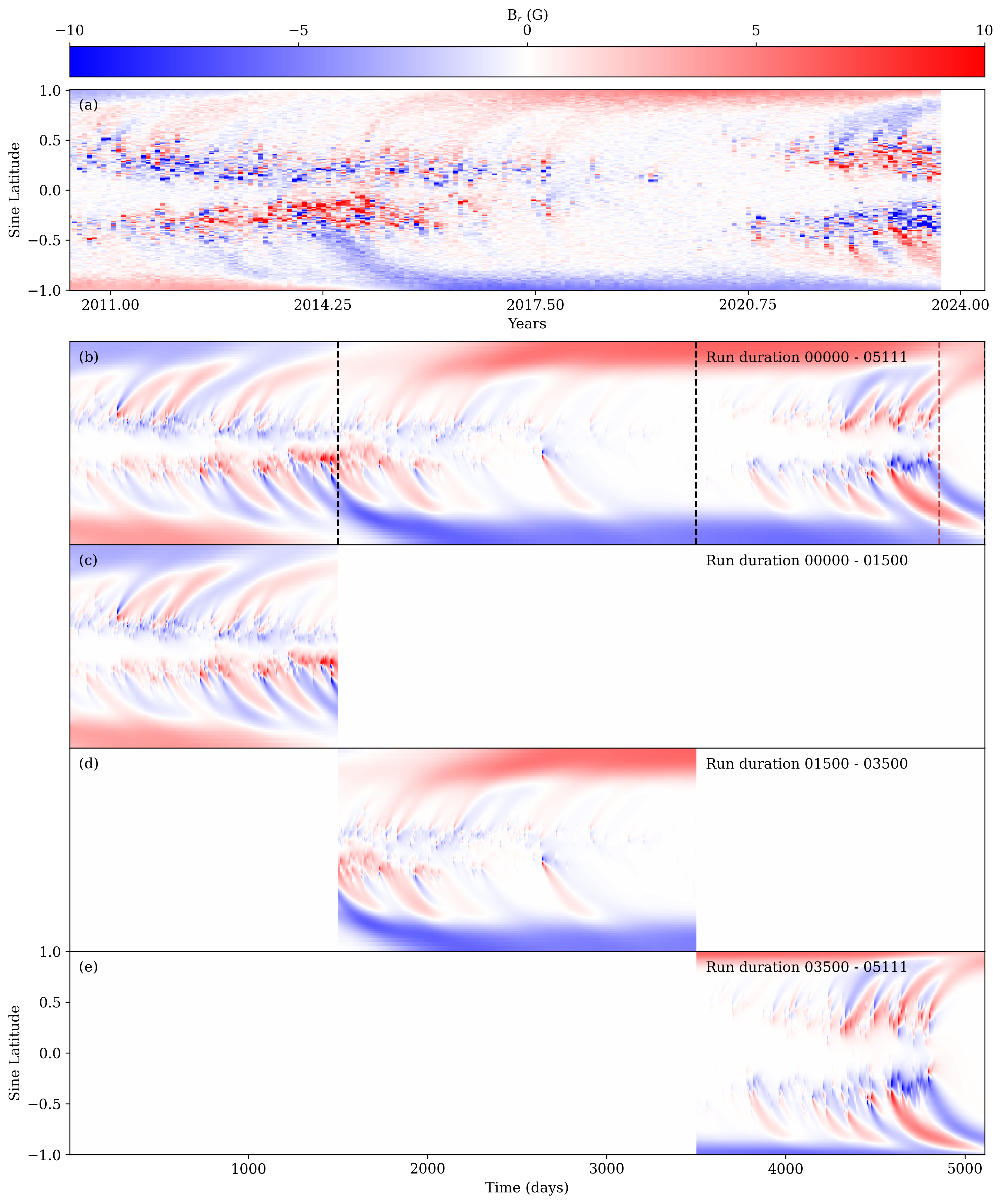}
     \caption{Butterfly diagram of surface radial magnetic field from observations and the SFT model, along with demonstrations of the ``restart'' functionality. Panel~(a) shows the HMI observed butterfly diagram interpolated onto SFT model resolution. A total of 14 years of SFT model evolution is plotted in~(b), where the brown dashed line toward the right indicates the time beyond which no new BMRs are incorporated into the SFT model.  In all the panels magnetic field saturates at $\pm$10G. Panel~(b) shows the uninterrupted SFT run for 14 years, where the black dashed lines indicate the time stamps at which the computation is stopped and restarted, the results of which are shown below. Panel (c) shows SFT evolution from $t$=0 day to $t$=1500 days. Panel (d) shows SFT evolution from $t$=1500 days to $t$=3500 days. Panel (e) shows SFT evolution from $t$=3500 days to $t$=5111 days. EnKF data assimilation requires the numerical model to have restart functionality.}
    \label{fig:2}
\end{figure*}

We evolve the SFT model for 14 years to generate a butterfly diagram of $\overline B(s,t)$. The initial condition for our simulation is the magnetic field distribution for Carrington map 2097. For our experiment, we used $\eta$=450~km$^2$~s$^{-1}$ and $v_0$=30.7~m~s$^{-1}$ based on several trial runs, as these values best reproduce the observed polar fields. An analogous butterfly diagram of $B_r$ is plotted from HMI Carrington maps for comparison. Figure~\ref{fig:2} shows the comparison between HMI observed radial component of the surface magnetic field and SFT generated butterfly diagram of surface radial magnetic field. The brown dashed line toward the right of Figure~\ref{fig:2}(b) indicates the last BMR insertion into the SFT model. Beyond this line, the model only shows the forward run of SFT model without new active regions.

It is important to note that for our DA experiment, the numerical model must have ``restart'' functionality. This is because the data assimilation modifies the model parameter (peak meridional flow speed for our case) constrained by the surface magnetic field observations at a user-defined assimilation cadence. After each modification, the simulation is restarted with the new value of peak meridional flow speed for the next iteration of assimilation window. We model the restart functionality in our simulation and compare the resulting magnetic field at random time frame to ensure the correctness of magnetic fields at the last frame and continuation in the next frame. Figure~\ref{fig:2}(b) denotes the uninterrupted SFT evolution for 14 years (5111 days, considering the leap years). Panel (c), (d) and (e) show partial SFT model evolution starting from 0 day, 1500 days, and 3500 days respectively. 

In order to check the magnetic field at the end of an SFT run and the start of the next instance, the latitudinal variation of $B_r$ is independently plotted and compared with the corresponding time step of panel (b). The continuity in the butterfly diagram in panel (c), (d) and (e) also indicate the evolution of surface magnetic field distribution. Magnetic field profile and amplitude at these iterations do not show a significant difference when compared with corresponding time step of panel (b) which validates the restart functionality of our model. Now the numerical model is ready to be coupled with the EnKF algorithm to infer various model parameters.

\subsection{Coupling the Ensemble Kalman filter algorithm with the Surface Flux Transport model}\label{subsec:enkf}


Data assimilation is performed by the EnKF as implemented in Data Assimilation Research Testbed \citep[DART,][\url{https://dart.ucar.edu}]{Anderson2009BAMS}. The algorithm operates through a two-phase process comprising a prediction phase and an update phase. During the prediction phase, the Kalman filter generates estimates of the current state variables along with their associated uncertainties. Then for the update phase these parameter estimates are used to compute the observed quantities and the corresponding uncertainties. In an iterative process, state variables are `nudged' closer to the true state that is inspired by the observations. An Observing System Simulation Experiment (OSSE) designed with EnKF in this context helps to estimate the potential value of state variables guided by the simulated observations.

Here, we provide a concise description of the Kalman Filter algorithm starting from a Bayesian formulation, as it applies to our DA experiment. For a detailed and more generalized mathematical derivation, readers can refer to \cite{Dikpati2016ApJ}. The surface flux transport process can be numerically modeled by solving the equations of the SFT model (Eq.~\ref{eq:2}) that require prescription of meridional flow (Eq.~\ref{eq:3}). One of the parameters in the presumed flow profile is the maximum flow speed in both hemispheres, which can be considered as our state variable ($D_u$ in Eq.~\ref{eq:3}). One can use $x_i$ for the range of state variables involved in modeling a physical system, where $i$ is the index of variable type and location in range $[1,n]$. Here $n$ denotes the dimension of state space. Ideally, the state vector ($\mathbf{x}_t$) can evolve in a time-dependent fashion that can be expressed as
\begin{equation}
\label{eq:4}
    \mathbf{x}_t = \mathbb{M}(\mathbf{x}_{t-1},t).
\end{equation}
In data assimilation terminology, the physical SFT model is the forward operator, whereas $\mathbb{M}$ is the ``model", which is a random-walk operator that generates the prior flow-speed from prior time-step t-1. In our case, the state vector $\mathbf{x}_t$ is the peak meridional flow speed ($D_u$), which is time-independent.

For our experiment we assume a prescribed meridional flow profile (Eq.~\ref{eq:3}) that does not vary with time. Using the vector notation, we can describe the state vector $\mathbf{x} \equiv x^i$ and the corresponding observations $\mathbf{y} \equiv y^j$ where the index $i$ denotes the state vector space that has a range $[1,n]$ and the index $j$ shows the observations with a range $[1,m]$. As stated earlier, $n$ is the dimension of the state space, whereas $m$ is the dimension of the observation space. While reconstructing the time-independent flow-speed by assimilating data in the SFT model, the state vector dimension is 1, i.e. $n=1$ in our case. As per EnKF-DART requirements, the observation space dimension will have to be less than or equal to the dimension of state vector. Thus $m$ is also 1. In our OSSE, the state vector $x^i$ is inferred from the observations $y^j$ at different assimilation steps. A detailed description of the integration of SFT in EnKF framework is provided in \S\ref{subsec:linking}.

For an assimilation time window $m_t$, a set of observations $y^i_t$ are available that can be generated using OSSE. Usually $m_t$ can vary with time and is much smaller than $n$. The prior estimates of the set of observations can be generated by a forward operator $\mathbf{h}(\mathbf{x},t)$ as,

\begin{equation}
    \label{eq:5}
    \mathbf{y^0} = \mathbf{h}(\mathbf{x},t) + \mathbf{\epsilon}^0(\mathbf{x},t),
\end{equation}
where the forward operator ($\mathbf{h}$) and state variable ($\mathbf{x}$) are assumed to model the observations accurately and $\mathbf{\epsilon}^0$ is the uncertainty associated with the observations. For our DA experiment, SFT model equation (Eq~\ref{eq:2}) is the forward operator and meridional peak flow speed ($v_0$) is the state variable that can be expressed as Eq.~\ref{eq:3} which is analogous to Eq.~\ref{eq:4}. In our OSSE system, we choose the value of the magnetic field at 60$^{\circ}$N latitude to be the observations ($\mathbf{y^0}$).

The uncertainty associated with the simulated system may be due to (i) limitations of the forward operator or (ii) error in the initial conditions. In EnKF experiment, we can only estimate the type (ii) uncertainty. In the context of our SFT model, the model equations are averaged over longitude. Hence, the impact of differential rotation on the flux transport process is not modeled. We also assume that the magnetic flux is only advected in response to the large-scale flow properties. Such assumptions limit us from modeling the small-scale flux distribution and the EnKF algorithm will not be able to model the uncertainty associated with such assumptions. On the other hand we can estimate the errors associated with the initial condition of the model i.e., a synoptic map of magnetic field distribution and properties of the BMRs used to generate the source terms from the distribution of $\mathbf{x}$ in the ensemble. If $\mathbf{x}^{Tr}$ is the true state i.e., the true state variable, and $\mathbf{x}^{p}$ is the prior estimate of the state variable with EnKF (Eq \ref{eq:4}), then the error in the prior estimate ($\epsilon^p$) can be written as
\begin{equation}
\label{eq:6}
    \epsilon^p = \mathbf{x}^{p} - \mathbf{x}^{Tr}.
\end{equation}
Similarly, the posterior state $\mathbf{x}^{a}$ and the error ($\epsilon^a$) associated with it can be computed by,
\begin{equation}
\label{eq:7}
    \epsilon^a = \mathbf{x}^{a} - \mathbf{x}^{Tr}.
\end{equation}
Here the index $a$ denotes quantities of the analysis states i.e., the posterior states. In order to find the optimal $\mathbf{x}^{a}$, we require observation ($\mathbb{R}$) and model ($\mathbb{P}$) error covariance matrices,
\begin{eqnarray}
    \mathbb{R} = \overline{\epsilon^0 \epsilon^{0T}},\label{eq:8}\\
    \mathbb{P} = \overline{\epsilon^p \epsilon^{pT}},\label{eq:9}
\end{eqnarray}
where the bar denotes the statistical average and $T$ indicates the transpose of the corresponding column vector. We also evaluate the estimated analysis error covariance ($\mathbb{A}$),
\begin{equation}
\label{eq:10}
    \mathbb{A} = \overline{\epsilon^a \epsilon^{aT}}.
\end{equation}

Now with this system of prior and analysis variables, we pose the question: What would be the most likely state vector $\mathbf{x}^{a}$ and its estimated variance $\epsilon^{a}$, considering the prior state vector $\mathbf{x}^{p}$ and available observations $\mathbf{y}^{0}$? In probability notation $P(\mathbf{x}^{a}|\mathbf{x}^{p},\mathbf{y}^{0})$ is the probability of occurrence of $\mathbf{x}^{a}$ provided $\mathbf{x}^{p}$ and $\mathbf{y}^{0}$ are known. According to Bayes' theorem, the probability distribution of $\mathbf{x}$ given $\mathbf{y}^{0}$ is
\begin{equation}
\label{eq:11}
    P(\mathbf{x}|\mathbf{y}^{0}) = \frac{P(\mathbf{y}^0|\mathbf{x})P(\mathbf{x})}{P(\mathbf{y}^0)}.
\end{equation}
Assuming the state vectors ($\mathbf{x}$) are drawn from a multivariate Gaussian distribution, prior state $P(\mathbf{x})$ can be written as,

\begin{equation}
    \label{eq:12}
    P(\mathbf{x}) = \frac{1}{(2\pi)^{1/2}|\mathbb{P}^{-1}|}e^{-\frac{1}{2}|(\mathbf{x}^p - \mathbf{x})^T\mathbb{P}^{-1}(\mathbf{x}^p - \mathbf{x})|},
\end{equation}
and $P(\mathbf{y}^0|\mathbf{x})$ as,
\begin{equation}
    \label{eq:13}
    P(\mathbf{y}^0|\mathbf{x}) = \frac{1}{(2\pi)^{1/2}|\mathbb{R}^{-1}|}e^{-\frac{1}{2}|(\mathbf{y}^0 - \mathbf{h}(\mathbf{x}))^T\mathbb{R}^{-1}(\mathbf{y}^0 - \mathbf{h}(\mathbf{x}))|}.
\end{equation}
Substituting Eq.~\ref{eq:12} and Eq.~\ref{eq:13} in Eq.~\ref{eq:11}, we can reduce the problem into finding an optimum distribution of $\mathbf{x}$. By linearizing $\mathbf{h}$ around $\mathbf{x}^p$, we can calculate an extremum of $\mathbf{x}$ as,

\begin{equation}
    \label{eq:14}
    \mathbf{x}^a = \mathbf{x}^p + (\mathbb{P} + \mathbf{h}^T\mathbb{R}^{-1}\mathbf{h})^{-1}\mathbf{h}^T\mathbb{R}^{-1}[\mathbf{y}^0 - \mathbf{h}(\mathbf{x}^p)].
\end{equation}

This is the expression for the Kalman Filter. A more detailed mathematical description in a solar context is provided in \cite{Dikpati2016ApJ}. For EnKF, the posterior state/update is constructed by creating an ensemble distribution of state variable from a multivariate Gaussian distribution and using these prior states to compute the ensemble of posterior states. If the errors associated with the observations are not expected to be correlated at different time steps, we can employ the sequential Bayesian analysis to accelerate the computation of EnKF. It is worthy of note that sequential Bayesian analysis is a statistical method used for updating the probability estimate of a state-vector as new observations become available. After each update, the new posterior probability becomes the prior probability for the next iteration of analysis. This is known as the Ensemble Adjustment Kalman Filter (EAKF) proposed by \cite{Anderson2001MWRv}. We have used this variant of EnKF for our experiment. 

\subsubsection{Applying Ensemble Kalman Filter Data Assimilation to Surface Flux Transport Model} \label{subsec:linking}

In the EnKF framework, the state vector ($\mathbf{x}_t$ in Eq. \ref{eq:4}) prescribes the meridional flow in our SFT model where the peak flow speed (defined by Eq. \ref{eq:3}) is one of the model parameters. In the following sections, we will be performing DA experiments to infer the value of peak meridional flow speed $v_0$, which can be calculated from $D_u$ following \citet{Yeates2020SoPh},

\begin{equation}
    v_0 = \pm D_u p^{p/2}(1+p)^{-(1+p)/2}.
    \label{eq:du_v}
\end{equation}

\begin{figure}[htpb]
    \centering
    \includegraphics[width=\linewidth]{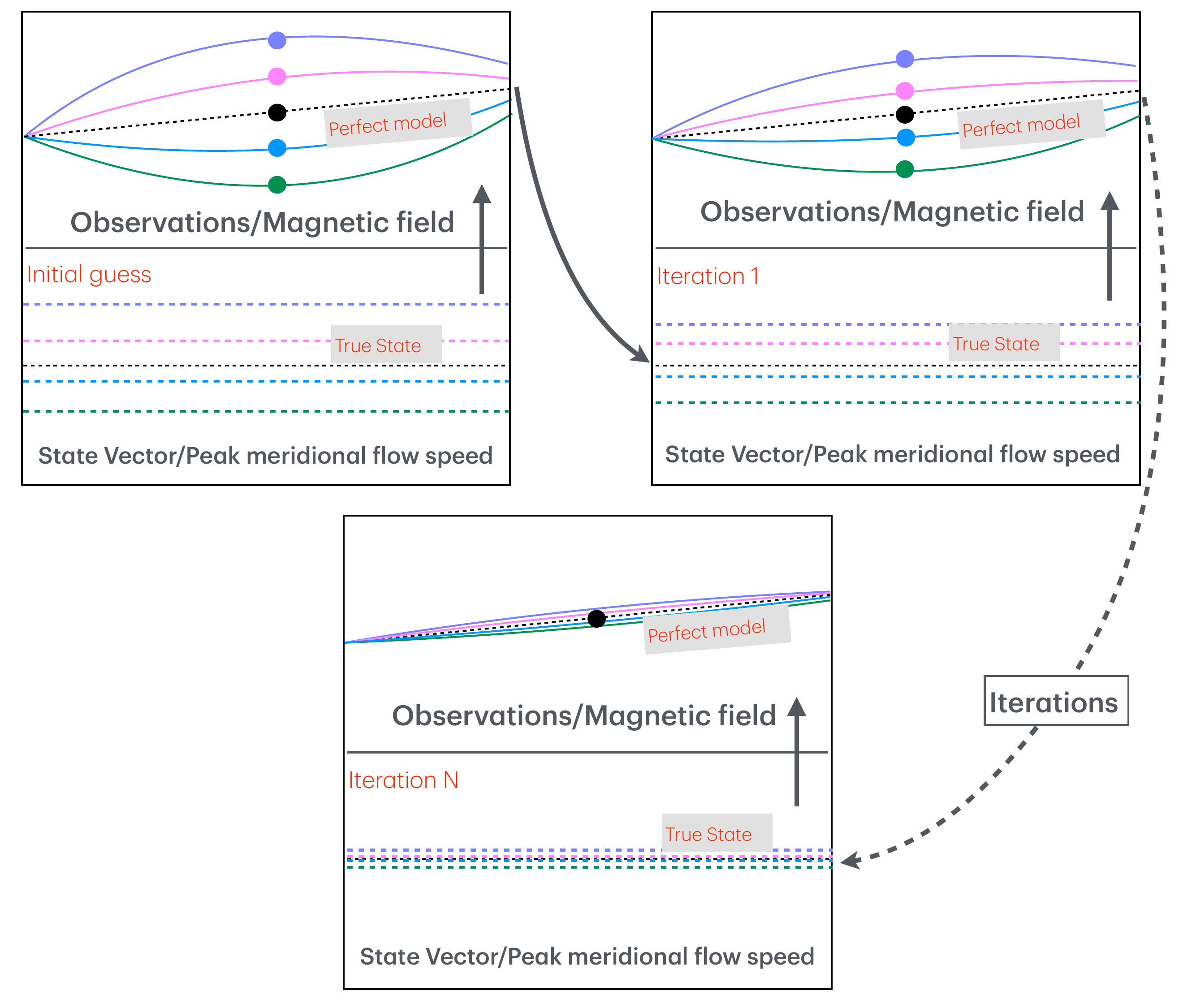}
    \caption{The illustration shows how the surface flux transport model is coupled to the Ensemble Kalman Filter data assimilation algorithm. Initially an ensemble of state vectors i.e., the peak meridional flow speed is generated and utilized in the SFT model evolution to calculate the corresponding observations i.e., the magnetic fields. In the next iteration, the state vectors were computed using the EnKF algorithm, where the state vectors were `nudged' towards the true-state. The process drives the resulting observations towards the perfect model. The iteration time step is decided by the assimilation cadence of the DA experiment. This iterative process is continued for the whole time duration of the global SFT calculations. e.g. if the global SFT simulation is set to evolve for 300 days, for an assimilation cadence of 30 days the iteration $N$ would be 100.}
    \label{fig:enkf_sft_algorithm}
\end{figure}

For $D_u$=0.085~km~s$^{-1}$ and $p$=2.33, the peak meridional flow speed is $v_0$=30.7~m~s$^{-1}$. With our state vector prescription, the observations ($\mathbf{y^0}$ in Eq. \ref{eq:5}), i.e., the magnetic field is computed using the forward operator ($\mathbf{h}(x,t)$ in Eq. \ref{eq:5}). A perfect SFT model is computed which provides the values of magnetic fields at all latitudes for the above flow parameters. In the data assimilation framework, an ensemble of initial guess for the state vectors is generated, and is used to evolve the SFT model to next iteration. Since the initial guesses for the state vectors are drawn from a multi-variate Gaussian distribution, the resulting observations do not necessarily generate identical observations as the perfect model. In the next iteration, the state vectors are adjusted according to the Bayes' theorem (Eq. \ref{eq:11}). This iterative process is continued until the end of the SFT evolution. In this iterative process the state vector is adjusted according to the observations as depicted by the illustration in Figure \ref{fig:enkf_sft_algorithm}.

\section{Results} \label{sec:results}

\begin{figure}[htpb]
    \centering
    \includegraphics[width=\linewidth]{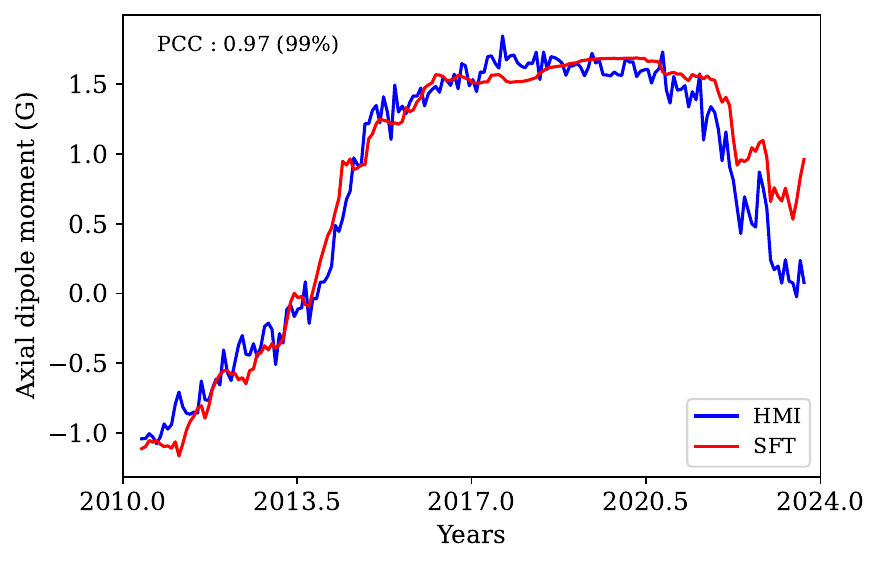}
    \caption{Axial dipole moment ($DM$) from HMI observations and the SFT model. The blue solid line denotes the DM time series calculated using HMI observations (butterfly diagram) and the red line indicates a similar calculation for the SFT modeled magnetic field. A Pearson correlation coefficient value of 0.97 between the HMI and SFT modeled axial dipole moment indicates that the numerical model performs very well.}
    \label{fig:5}
\end{figure}

Using the SFT model described in \S\ref{subsec:sft}, we evolve the surface radial magnetic field for about 14 years using a peak meridional flow speed $v_0=30.7$~m~s$^{-1}$. The resulting butterfly diagram is compared to the observations in Figure~\ref{fig:2}. In order to test our model output, we compare the calculated axial dipole moment ($DM$ in Eq.\ \ref{eq:dm}) with the observations in Figure~\ref{fig:5}. The Pearson correlation coefficient (PCC) between these two time series is 0.97, which indicates that the SFT model output is reasonable. For our current experiment we choose the value of magnetic field at 60$^{\circ}$N latitude, noting that the result of DA experiment does not depend on the location of the reference magnetic field value. In principle, one can choose to utilize the values of magnetic field at all latitudes as their true observations. However, this makes the setup computationally expensive due to increased dimensionality. 

The resulting butterfly diagram from the SFT model (Figure~\ref{fig:2}(b)) is used as our ``true observations'' for the data assimilation experiment with a true state (peak meridional flow speed) value of $v_0=30.7$~m~s$^{-1}$. In order to check the sensitivity and robustness of the assimilation results we explore variations in: (a) assimilation cadence, (b) error associated with the true observations, and (c) ensemble size. We define our perfect model with a peak meridional flow speed of $v_0=30.7$~m~s$^{-1}$. Depending upon the assimilation cadence, the perfect model observations (i.e., the magnetic fields at 60$^{\circ}$N latitude) for the corresponding time step from the original SFT model is considered for our DA experiment. For different experiments with different ensemble size and error variance, the perfect model calculations are performed with the same parameters. Only the magnetic field at 60$^{\circ}$N latitude generated by the perfect model is used for computing the state vectors for these different cases. Consequently, the state vectors are `nudged' based only on observations sampled at that latitude point. Under the current setup of EnKF data assimilation, each experimental case takes approximately 45~min to complete for an assimilation cadence of 60 days and an ensemble size of 20. Reconstruction of the state vector at multiple spatial locations for a larger ensemble size would be more computationally demanding. 

\subsection{Impact of different assimilation cadence on data assimilation}\label{subsec:assimilation_cadence}

Assimilation cadence is probably one of the most important parameters in a data assimilation experiment. This corresponds to a time after which the DA simulation is interrupted, the value of the state vector is compared with the true states, and a posterior state vector is created using Eq.~\ref{eq:14}. The ensemble size chosen for this study is 20 with an error variance value of $10^{-6}$. Our SFT model has a time step of $dt$=4~hr. We run our DA experiment for assimilation cadences of 1, 10, 20, 30, 60, 100, 120, 150, 180 and 200 days. The choice of these cadence values is arbitrary. For all model runs, the assimilation cadence denotes the time interval after which the state vector (i.e., the peak meridional flow speed) is modified and the observation (i.e., the magnetic field) is sampled.

\begin{figure*}[htpb]
    \centering
    \includegraphics[width=\linewidth]{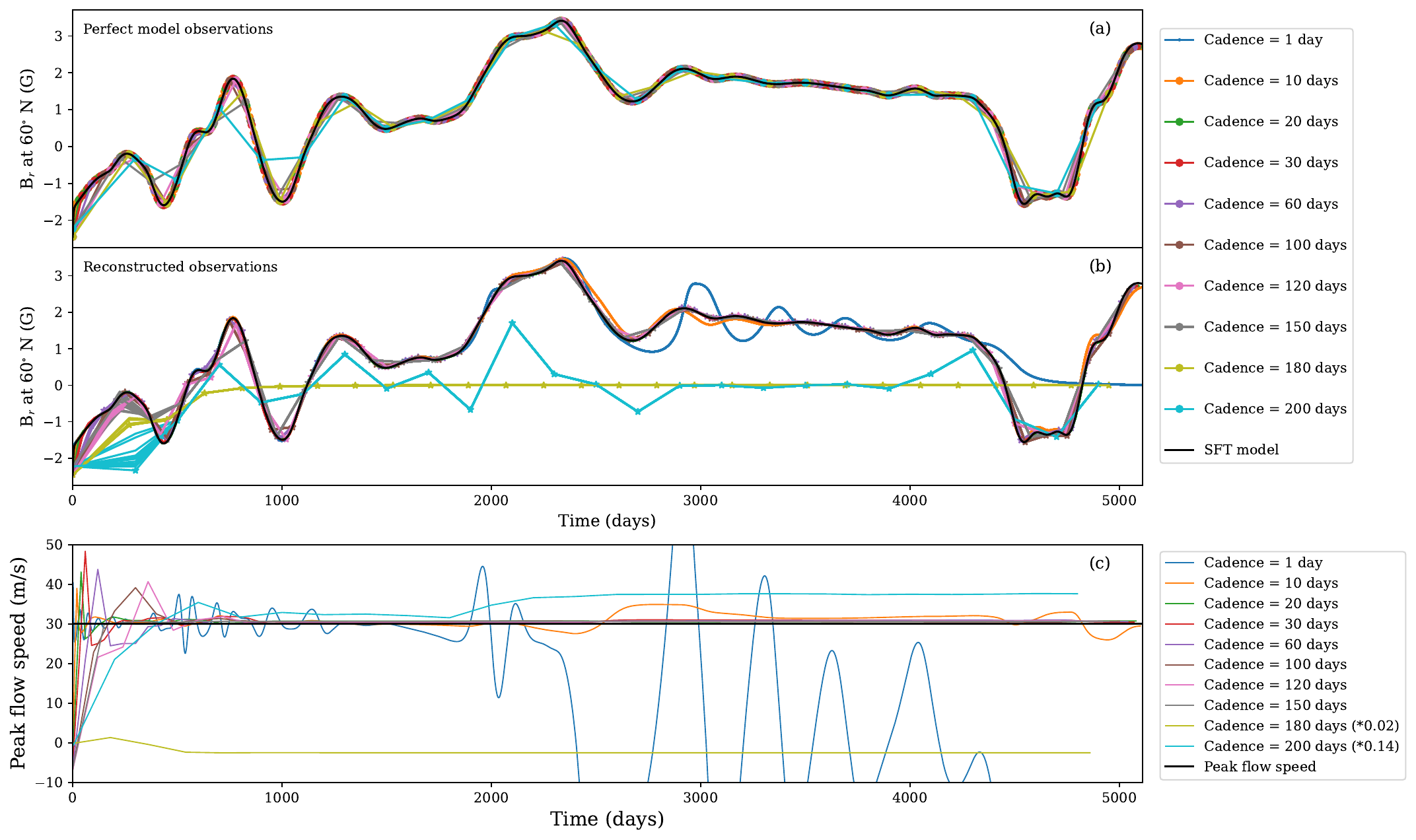}
    \caption{True observations, reconstructed observations and the state vectors are plotted for different assimilation cadences. (a) Perfect model observations (magnetic field at 60$^{\circ}$N latitude) for different assimilation cadence intervals. (b) Reconstructed magnetic field at 60$^{\circ}$N when meridional flow is iteratively `nudged' towards the true state vector (peak meridional flow speed). (c) Evolution of peak meridional flow speed (state vector) for these cases where the variable is approaching the parameter value of the perfect model calculation. The solid black line indicates the outputs from the original SFT model and the different colors correspond to various assimilation cadences. Note that for very high (more than 150 days) and very low (less than 10 days) assimilation cadences, the state vector values do not converge towards the true state. State vectors for assimilation cadences of 180 days and 200 days are re-scaled to fit in the plot.}
    \label{fig:6}
\end{figure*}

\begin{figure}[htpb]
    \centering
    \includegraphics[width=\linewidth]{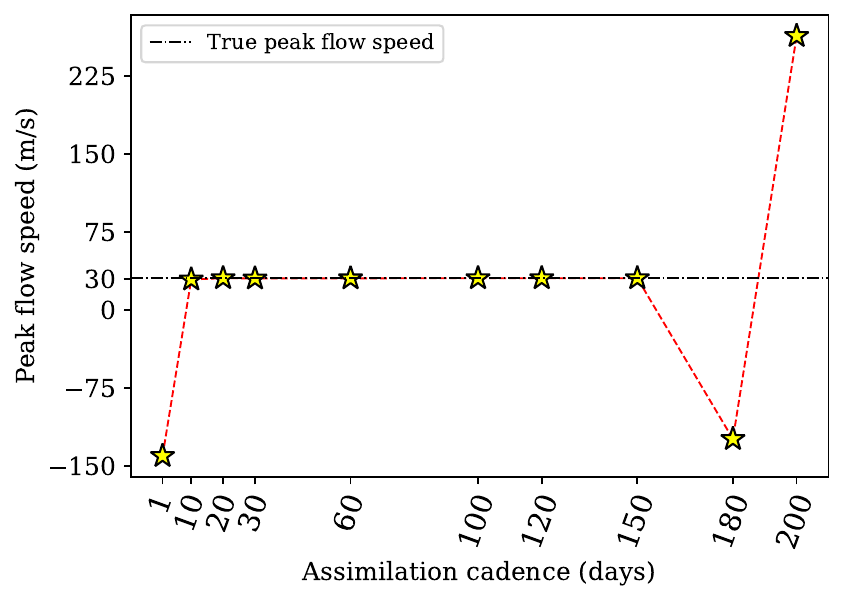}
    \caption{Ensemble mean of the state vector at the end of DA experiment for different assimilation cadence. The horizontal axis denotes the different cadence values and the vertical axis indicates the value of the ensemble mean of state vectors at the end of the DA experiment. It is essential to choose an appropriate value of data assimilation frequency in order to correctly reconstruct the observations. The black-dashed curve depicts the true state value for reference.}
    \label{fig:7}
\end{figure}

Figure~\ref{fig:6} (a) shows the true observations (value of the magnetic field from the SFT model) in different colors. Figure~\ref{fig:6} (b) indicates the reconstructed observations (value of the magnetic field from the SFT model after the DA). The initial deviation from the perfect model in this plot shows the impact of data assimilation experiment where the state vector (peak meridional flow speed) is iteratively `nudged' towards the perfect model. Figure~\ref{fig:6} (c)  shows the evolution of one of the state vectors (value of peak meridional flow speed) from the ensemble. For all the cases, the baseline value of magnetic field and $v_0$ is solid black. For some cases, we observe a deviation from the true observations. To summarize, we show the final mean value of the ensemble of state variables in Figure~\ref{fig:7}. Here the markers denote the ensemble mean peak meridional flow speed at the end of the DA experiment for different assimilation cadence instances. For assimilation cadences between 10 days to 150 days the reconstructed observations and corresponding peak meridional flow speed are close to the true state values. We discuss the possible cause of the spread from the true state for a few cases of high assimilation cadence value within the plot in \S\ref{subsec:discussion}.  

\begin{figure*}[htpb]
    \centering
    \includegraphics[width=\linewidth]{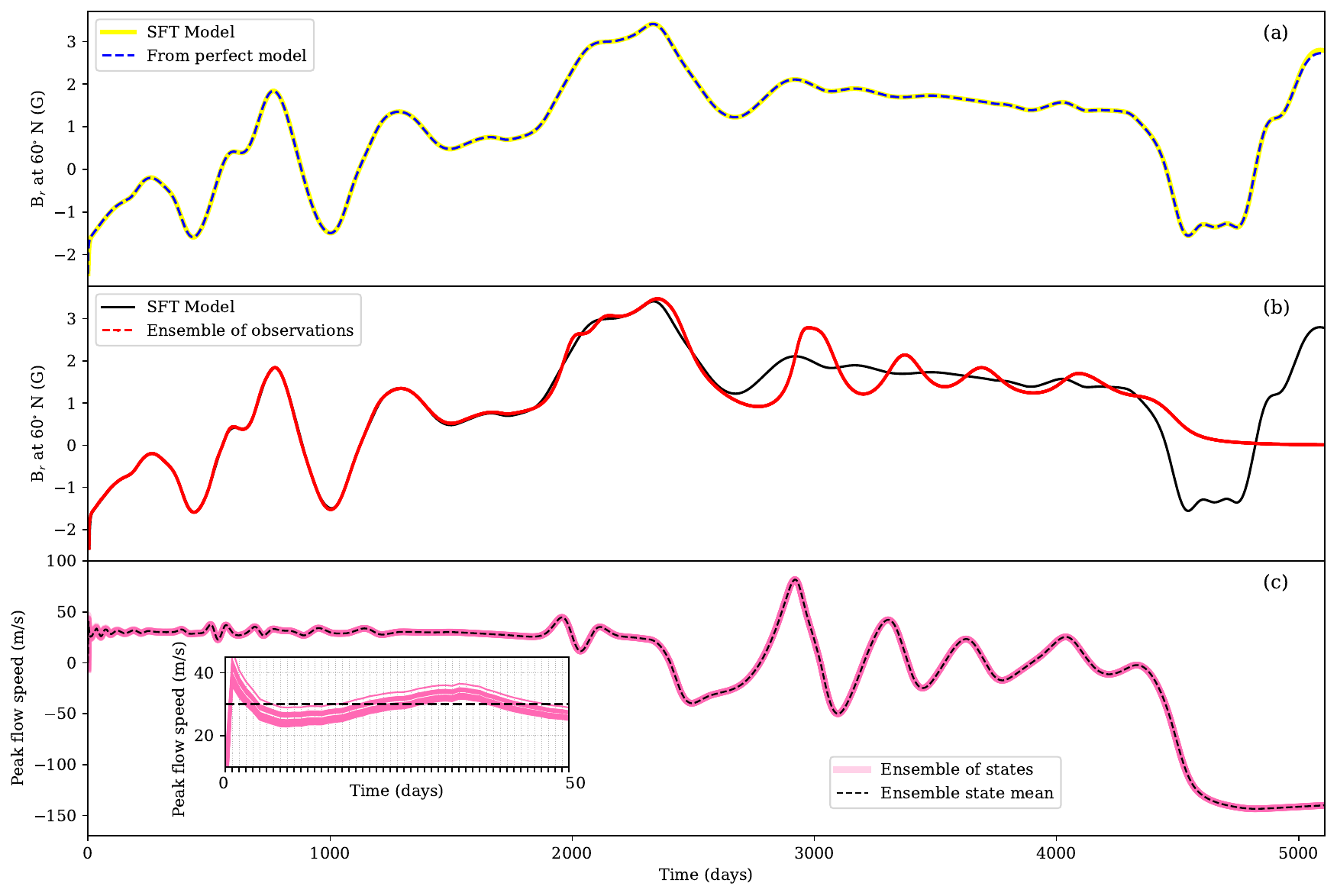}
    \caption{Data assimilation with an assimilation frequency of 1 day. True observations (magnetic field at 60$^{\circ}$N latitude ), reconstructed magnetic fields in red  solid line, and the variation in the peak meridional flow speed is plotted in panel (a), (b) and (c) respectively. Such a high assimilation cadence does not provide appropriate relaxation time to the forward operator, hence the resulting reconstruction deviates from the perfect model. The inset plot in (c) denotes the first 50 days of the assimilation process indicating the oscillatory behavior around the true state (dashed black curve) before approaching the true state. Such oscillations are found for all the cases of the DA experiment.}
    \label{fig:8}
\end{figure*}

To explore the reason for deviations at low and high assimilation cadences, we inspect the results for the case of 1 day, that is, we modify the ensemble of state vectors after each day. In Figure~\ref{fig:8} we plot the true observations (SFT magnetic field at 60$^{\circ}$N latitude), reconstructed observations, and state variable (peak meridional flow speed) in panel (a), (b) and (c) respectively. Initially the state variable and the reconstructed observations are oscillating around the true state; however, at later times they diverge drastically. It is apparent that for such a high cadence of assimilation the reconstructed observations are not converging towards the true value at the end of the simulation. This is commonly known as ``ensemble collapse", which is discussed in \S\ref{subsec:discussion}. 

\subsection{Impact of different error variance on data assimilation}\label{subsec:error_variance}

\begin{figure*}[htpb]
    \centering
    \includegraphics[width=0.93\linewidth]{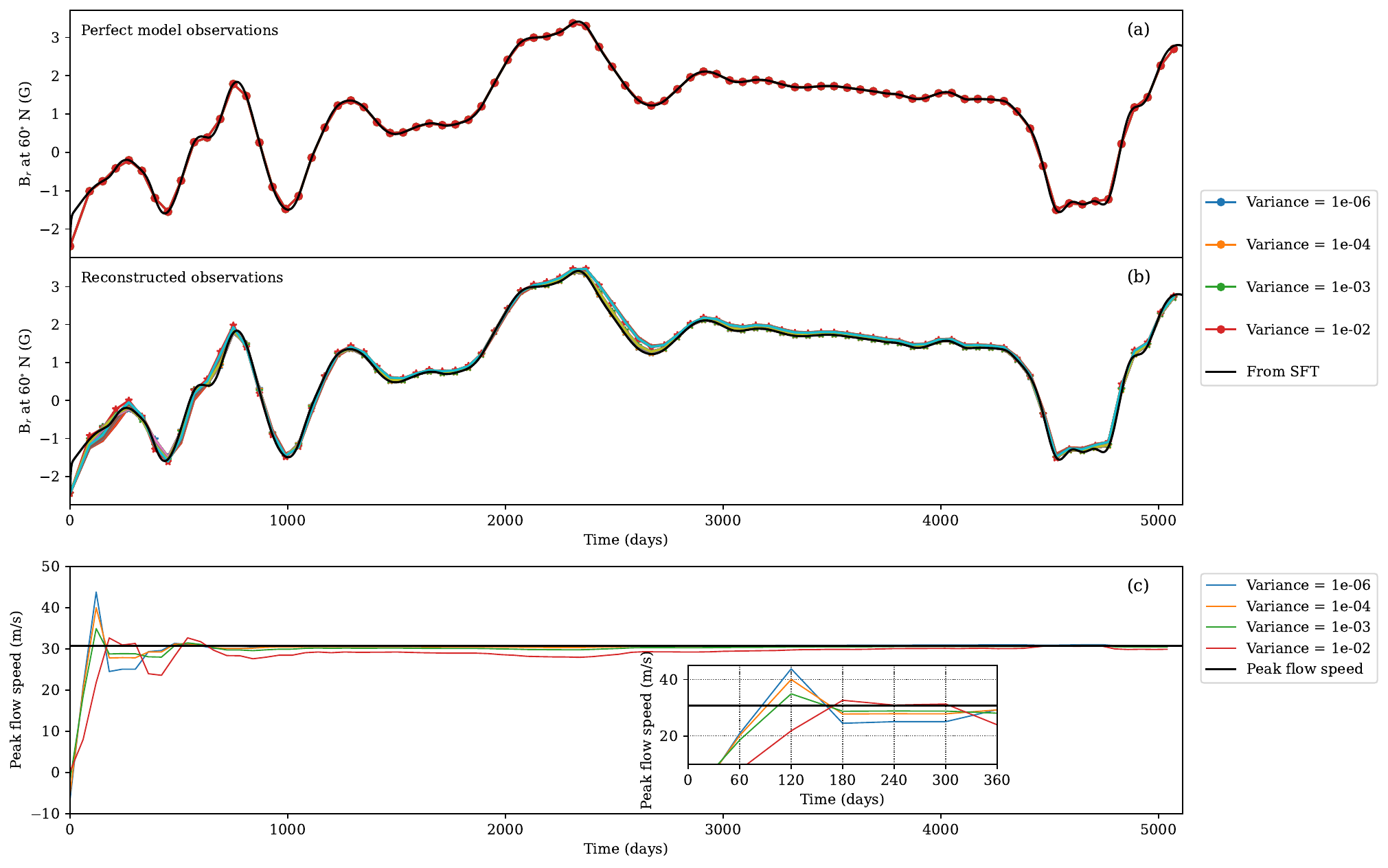}
    \caption{Probable variation in the error associated with the perfect model and its implications for the reconstruction process. We have chosen an assimilation cadence of 60 days and ensemble size of 20 for this experiment. Panel (a) shows the perfect model observations i.e., magnetic fields at 60$^{\circ}$N latitude for different error variance values while (b) denotes the reconstructed observations. In panel (c) we plot the corresponding state vectors for different error variance values. We notice a deviation in the reconstruction from the perfect model when the associated error variance is higher. The inset plot highlights the initial oscillatory pattern of the state vectors for different experimental cases before converging towards the true state (denoted by the solid black curve).}
    \label{fig:9}
\end{figure*}

\begin{figure}[htpb]
    \centering
    \includegraphics[width=\linewidth]{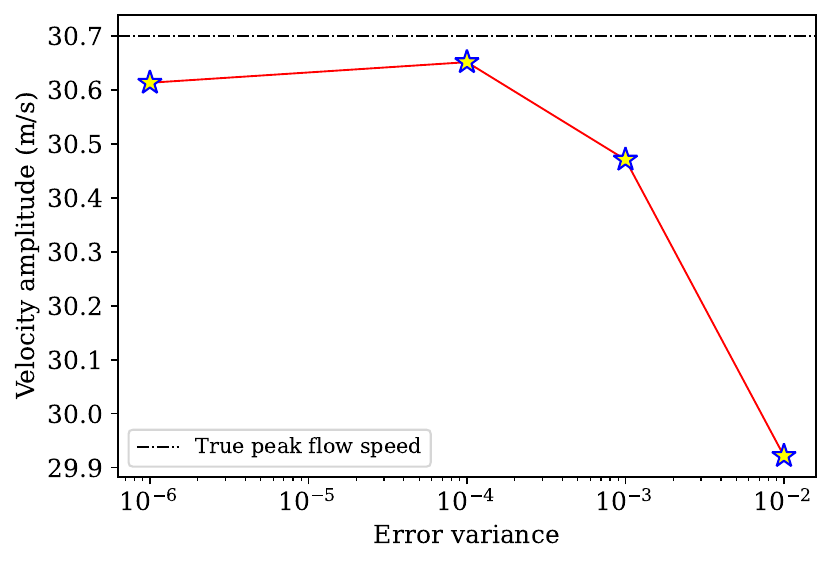}
    \caption{Ensemble mean value of the state vector at the end of the DA experiment for different error variances. The assimilation cadence is chosen to be 60 days and ensemble size is 20 for this experiment. DA experiment is sensitive to the degree of error associated with the perfect model. The black-dashed curve shows the true state value for reference.}
    \label{fig:10}
\end{figure}

Uncertainty associated with modeled surface magnetic field may originate from the initial conditions or accuracy of the numerical schemes. In our case, the output of the SFT model is an ``exact" numerical solution of a partial differential equation within the limit of our numerical discretization scheme, which does not have any uncertainties. However, for the case of direct magnetic field observations, there can be associated uncertainties. Therefore we explore a parameter space of different error variance associated with the true observations. For this study, we choose an assimilation cadence of 60 days and ensemble size of 20. We run our DA experiment for relative error variance values of 10$^{-6}$, 10$^{-4}$, 10$^{-3}$ and 10$^{-2}$. Figure~\ref{fig:9} shows a comparison between the perfect model observations and the reconstructed observations along with the state vectors for different cases of error variance value. For most of the cases, the reconstruction is reasonably good with minor deviation from the perfect model. Since the ensemble of state vectors for the assimilation process are drawn from a multivariate Gaussian distribution, the algorithm shows an oscillatory pattern around the true state before finally converging, as highlighted by the inset plot in Figure~\ref{fig:9} (c). In Figure~\ref{fig:10} we plot the final ensemble mean value of the state vector at the end of the DA experiment. In all cases, the inferred meridional peak flow speed from our DA experiment is close to the true state value; the maximum difference is less than $1$~m~s$^{-1}$. The assimilation cadence chosen here is 60 days. We intentionally choose this assimilation cadence to highlight the fact that even though there are fewer observations available to assimilate, the degree of error associated with those observations can modify the inferred state vector. 

\subsection{Impact of different ensemble size on data assimilation}\label{subsec:ens_size}

\begin{figure*}[htpb]
    \centering
    \includegraphics[width=0.92\linewidth]{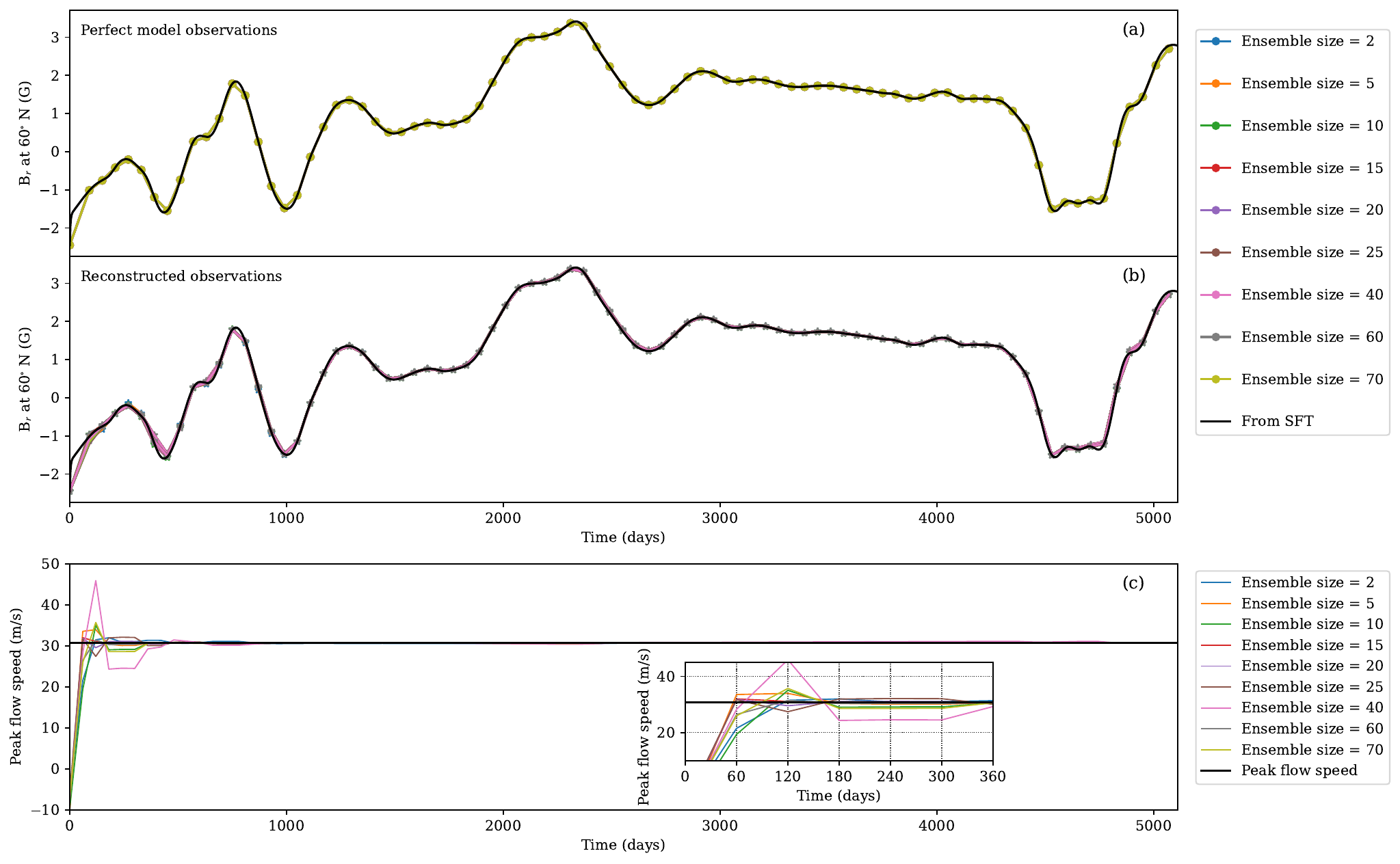}
    \caption{Reconstruction using different ensemble sizes. We have chosen an assimilation cadence of 60 days and error variance of 10$^{-6}$ for this experiment. (a) Perfect model observations i.e., magnetic field at 60$^{\circ}$N latitude with varying ensemble sizes. (b) The reconstructed magnetic fields after the DA experiment for respective ensemble sizes. (c) The corresponding evolution of state vectors i.e., the peak meridional flow speed as they approach towards the true state. Although the assimilation cadence is 60 days, increasing ensemble size improves the reconstructed observations. In the inset plot shown in the bottom panel, we plot the initial iterations of the DA experiment showing the oscillatory pattern of the state vectors around the true state (denoted by the solid black curve).}
    \label{fig:11}
\end{figure*}

\begin{figure}[htpb]
    \centering
    \includegraphics[width=\linewidth]{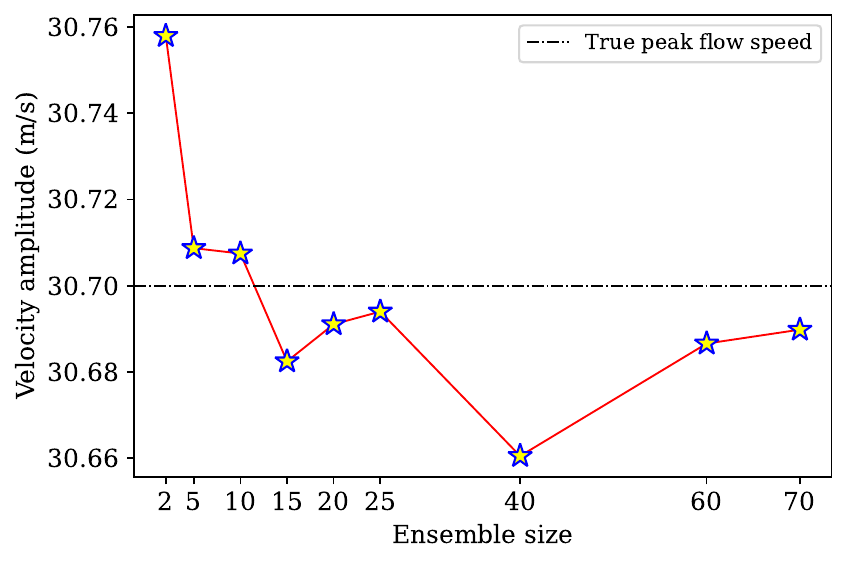}
    \caption{Ensemble mean value of the state vector at the end of the DA experiment for different ensemble sizes. The assimilation cadence is chosen to be 60 days and error variance of 10$^{-6}$ for this experiment. The final state vector approaches the true state for a larger ensemble size even for a higher assimilation cadence. The black-dashed curve denotes the true state value for reference.}
    \label{fig:12}
\end{figure}

In our data assimilation scheme, the size of the ensemble is one of the controlling parameters that can impact the robustness of the assimilation process. On one hand, in order to produce a meaningful reconstruction the size should not be too small. On the other hand, the size of the ensemble plays a key role in effective use of computational resources, as the forward operator has to compute the observations for a given state vector. \citet{Anderson2010MWRv} shows that choosing a larger ensemble size does not necessarily improve the assimilation results. For our DA experiment, we explored the impact of different ensemble sizes e.g., 2, 5, 10, 15, 20, 25, 40, 60 and 70. Figure~\ref{fig:11} shows a comparison between the perfect model and the reconstructed observations along with the state vectors for computations with different ensemble sizes. We have chosen an assimilation cadence of 60 days and error variance of 10$^{-6}$ for this study. For all the cases, initially the algorithm produces state vectors that oscillate around the true state before converging with minor deviations from the perfect model. In Figure~\ref{fig:12} we plot the final ensemble mean value of the state vector at the end of the DA experiment. Here the black dashed curve denotes the true state for reference. As we increase the ensemble size the assimilation result seems to improve while saturating towards the end. In our DA experiment, we have prior knowledge of the true state parameter (i.e., the peak meridional flow speed), however when such an assimilation technique is applied to observations of magnetic field to extract the peak meridional flow speed, the choice of a suitable ensemble size is crucial for meaningful interpretation of the results. The difference in the state vector for ensemble size 40 is only ~0.04 ~m~s$^{-1}$ which does not significantly alter the modeled magnetic field. However, we believe that for an assimilation process involving multi-dimensional spatial reconstruction of global flow characteristics, such degree of sensitivity of the algorithm will be helpful.

\section{Discussion and Conclusion} \label{subsec:discussion}

In this study, we utilized the SFT model to infer the peak meridional flow speed within a data assimilation framework of EnKF. For this purpose we developed a longitudinally averaged SFT model experiment and evolved the model for $\approx$~14~yr. In order to check the accuracy of the SFT evolution, we compared the axial dipole moment from our model to the HMI observations. We noticed that adjusting the DA parameters e.g., assimilation cadence, the error associated with the perfect model can modify the reconstruction of the observations. 

Direct observation of the photospheric magnetic field has its challenges (e.g., for high latitudes and the far side), but these measurements are essential for driving global coronal models \citep{Linker1999JGR,Schrijver2003ApJ,Schrijver2013ApJ,vanderHolst2014ApJ,Hoeksema2020ApJS}, solar wind simulations \citep{Owens2017NatSR}, reconstruction of three-dimensional inner heliospheric states \citep{Owens2013LRSP}, and space weather applications \citep{Arge2000JGR}. The polar fields are difficult to measure accurately, but are particularly important for solar cycle prediction studies \citep{Jiang2010ApJ, Upton2014ApJ,Bhowmik2018NatCo}. Global surface-flow characteristics play an important role in modulating the polar fields. This warrants the usage of SFT simulations constrained with available magnetic field observations -- utilizing the theoretically modeled flow distribution that best matches the measurements -- to produce true global photospheric magnetic field distributions that include polar field estimates. Apart from the global flow characteristics, the active region flux and the mid-latitude poleward flow speed show a negative correlation, which might be attributed to the field-dependent converging flow towards flux concentrations \citep{Sun2015ApJ}. This firmly establishes the requirement for techniques to infer the flow characteristics from observational proxies.

Utilizing our EnKF + SFT method, we are able to constrain the surface flow properties with magnetic field measurements, which will improve the current global surface flux distribution modeling efforts. Therefore, we built an OSSE setup where one of the flow parameters, the peak meridional flow speed, is estimated from synthetic observations of magnetic fields. 

Several control parameters within the assimilation framework can be adjusted to improve the assimilation process. When such an algorithm is applied to a system of observed magnetic fields, there are constraints on the observation cadence (e.g., Carrington maps of magnetic fields are only available at a cadence of ~28 days) and sometimes there are data gaps from the observing stations. With the EnKF setup we can assimilate available observations having different or irregular cadence without compromising the reconstruction process. It is important to note that the forward operator (SFT model in our case) requires sufficient time to relax. Hence, if the flow properties are modified within a shorter timescale (e.g. each day) it will lead to inconsistent results. This is known as ensemble collapse as illustrated by Figure~\ref{fig:8}. Hence the optimal assimilation cadence varies depending upon the properties of the forward operator. Apart from the assimilation cadence, we can also control the reconstruction process for different degrees of error associated with the observations. The algorithm also provides further constraints in terms of ensemble size for computing the state vector. In our experiment, EnKF algorithm reconstructs the state vector within first few months of simulation where the impact of the choice of ensemble size on the assimilation process is amplified. The final vales of the state vector corresponding to different experiment cases show negligible deviation from the true state. However, the SFT generated magnetic fields for different such choices still closely tracks the perfect model output. We believe, for a global flow reconstruction, such minor changes (e.g., for the case of ensemble size 40 in Figure~\ref{fig:12}) may impact the modeling of magnetic field distribution locally. 

Provided the versatility of the EnKF technique, usage of the same within the solar physics community has not been widely explored. These tools have been employed by various authors in a few contexts, such as solar cycle prediction, parameter reconstruction, and estimating solar wind properties \citep{Kitiashvili2016ApJ,Dikpati2016ApJ,Lang2017SpWea}.
However, a golden era for data assimilation in solar models is coming soon. In a recent ground-breaking effort, \cite{Turner2023SpWea} demonstrated that EnKF assimilation of near-real-time solar wind data into the WSA (Wang-Sheeley-Arge) model can improve the 5-day lead-time forecast by 15\% if L5 data is included in assimilation along with L1 data. 

The versatility and applicability of the EnKF technique across a range of scientific and practical problems demonstrates its effectiveness and calls for ongoing development in these areas. Nevertheless, the method does have several limitations. The EnKF assumes that the distributions of the forecast and observational errors are Gaussian, which can be a significant limitation when dealing with non-Gaussian processes. With a limited ensemble size, the EnKF can suffer from sampling errors, which can lead to inaccurate estimates of the state and error covariances. To improve the accuracy, large ensemble size can be used, but then the EnKF can be computationally expensive, especially for large-scale systems, because it requires the integration of a large number of ensemble members through the model. Two assumptions in the EnKF tool, namely (i) the model is ``perfect" (i.e., all physics are sufficiently included to reproduce the observations) and (ii) the covariance coefficients of the state vectors are considered to be linearly independent, may lead to inaccuracies in a highly nonlinear system \citep{Carrassi2018WIRCC}. 

In this paper we demonstrated the successful implementation of the EnKF algorithm with an SFT model with parameterized meridional flow profile. In future work we plan to explore the usage of our technique to infer two dimensional spatiotemporal descriptions of the global flows on the photosphere inferred from the magnetic field observations.

\begin{acknowledgments}
This work is supported by NASA LWS award 80NSSC20K0184 to Stanford University and a subaward to the University of Hawai`i. SD and XS additionally acknowledge support from the State of Hawai`i. This work is also supported by the NSF National Center for Atmospheric Research, which is a major facility sponsored by the National Science Foundation under cooperative agreement 1852977. MD acknowledges partial support from various NASA grants, such as NASA-LWS award 80NSSC20K0355 to NCAR and NASA-HSR award 80NSSC21K1676 to NCAR. SSM, YL, MLD and JTH were also partly supported by the COFFIES: NASA-DRIVE Science Center award 80NSSC22M0162. SSM, YL and JTH also acknowledge support from NASA contract NAS5-02139 (HMI) to Stanford University. Data used in this work are courtesy of NASA/SDO and the HMI science teams.
\end{acknowledgments}

\bibliography{references}{}
\bibliographystyle{aasjournal}

\listofchanges

\end{CJK*}
\end{document}